\documentclass[aps,showpacs,amsmath,amssymb]{revtex4}

\usepackage[dvips]{graphicx}

\newcommand{\bra}[1]{\mbox{$\left\langle{#1}\right|$}}
\newcommand{\ket}[1]{\mbox{$\left|{#1}\right\rangle$}}

\begin{document}

\title{Local implementation of nonlocal operations with block forms}
\author{Ning Bo Zhao\footnote{nbzhao@mail.ustc.edu.cn}, An Min Wang\footnote{anmwang@ustc.edu.cn}}
\affiliation{Quantum Theory Group, Department of Modern Physics\\
University of Science and Technology of China, Hefei 230026, People
Republic of China}

\begin{abstract}
We investigate the local implementation of nonlocal operations with the block matrix form, and propose a protocol for any diagonal or offdiagonal block operation.
This method can be directly generalized to the two-party multiqubit case and the multiparty case.
Especially, in the multiparty cases, any diagonal block operation can be locally implemented using the same resources as the multiparty control-U operation discussed in Ref. \cite{p00}.
Although in the bipartite case, this kind of operations can be transformed to control-U operation using local operations, these transformations are impossible in the multiparty cases.
We also compare the local implementation of nonlocal block operations with the remote implementation of local operations \cite{hpv01}, and point out a relation between them.
\end{abstract}


\maketitle
\section{Introduction}
Nonlocal operations are critical in distributed quantum computation.
Sometimes, a collective operation needs to be implemented on the qubits at distant nodes.
Generally, such an operation can not be implemented directly.
However, it is possible to be implemented locally, i.e., it can be implemented using local operations and classical communications (LOCC), shared entanglement and some auxiliary qubits.
Obviously straightforward method to implement such a nonlocal operation is using quantum state teleportations \cite{six93}, i.e., teleporting all of the qubits to one node, performing the operation at this node and teleporting these qubits back.

In the bipartite case, it requires two rounds of state teleportation and one local collective operation to implement a nonlocal operation using the above method.
So this process consumes two ebits (shared entanglement resources) and four cbits (classical communications).
These resources are necessary for some operations such as the SWAP operation \cite{p00,clp01}.
However, there are operations that can be locally implemented using less resources \cite{p00,clp01}.
For the CNOT operation, the necessary and sufficient resources are one ebit plus two cbits --- one cbit for each direction.
These resources are also sufficient for general control-U operations.
Ref. \cite{p00} presented a protocol to locally implement the CNOT operation just using these resources.
This protocol has been experimentally demonstrated in Ref. \cite{guo04}.
Ref. \cite{p00} also pointed out that a similar protocol --- replacing one CNOT operation by the control-U operation --- can be used for general control-U operation.

In Sec. \ref{p2}, we propose a similar protocol of local implementation of nonlocal operations with diagonal or offdiagonal block forms, using the same resources.
This protocol is free of the specific content of the blocks, so it is available even if these blocks are unknown.
This protocol is also independent of the dimension of the blocks, so it can also be used in the case if there are multiqubits at the node where the operation is actually implemented.
We compare the local implementation of nonlocal block operations with the remote implementation of local operations in Sec. \ref{pnm}.
We generalize the protocol to the multiqubit cases in Sec. \ref{pnm} and to the multiparty cases in Sec. \ref{p3}.
In Sec. \ref{con}, we summarize our results.

Recently, the problems of constructing a nonlocal operation or simulating it by other operations are discussed \cite{d1,d2,d3,guo05}.
We do not discuss this problem in this paper.
The problem discussed in this paper is how to locally implement a nonlocal operation using LOCC and shared entanglement resources if the device of the operation has been constructed in one node.

\section{nonlocal block operations on two qubits} \label{p2}
Consider these nonlocal diagonal block operations
\begin{equation}\label{u}
U =
\left( \begin{array}{cc}
u_0 & 0\\
0 & u_1
\end{array} \right),
\end{equation}
where $u_0$ and $u_1$ are $2\times 2$ unitary matrices.
Alice and Bob need to implement such an operation on their qubits \textit{A} and \textit{B}, where qubit \textit{A} belongs to Alice and qubit \textit{B} belongs to Bob.
Bob has the device to implement this operation.
$U$ in Eq. \ref{u} can be expressed as
\begin{equation}
U^{A,B}=\sum_{i=0}^{1} \ket{i}_A\bra{i} \otimes u_i^B.
\end{equation}
Let us propose the following protocol in order to locally implement such an operation on the qubits \textit{A} and \textit{B}.

In general, the jointed initial state of the qubits \textit{A} and \textit{B} can be expressed as
\begin{equation}
\ket{\Psi_0}_{AB}=\alpha_0 \ket{0}_A \ket{\xi_0}_B +\alpha_1 \ket{1}_A \ket{\xi_1}_B, 
\end{equation}
where $\ket{\xi_0}$ and $\ket{\xi_1}$ are arbitrary state and need not be orthogonal.

They share a maximally entangled pair $A_1B_1$ in the state
\begin{equation}\label{e}
\ket{\Phi}=\frac{1}{\sqrt{2}}(\ket{00}+\ket{11}),
\end{equation}
where qubit $A_1$ belongs to Alice and qubit $B_1$ belongs to Bob.

\paragraph*{step 1}
Alice performs a CNOT operation on her qubits $A$ and $A_1$, using the qubit $A$ as the control.
After this, the state of $A,B,A_1,B_1$ becomes
\begin{eqnarray}
& & CNOT^{A,A_1} \ket{\Psi_0}_{AB} \otimes \ket{\Phi}_{A_1B_1} \nonumber \\
&=& CNOT^{A,A_1} \sum_i \alpha_i \ket{i}_A \ket{\xi_i}_B \frac{1}{\sqrt{2}} \sum_j \ket{jj}_{A_1B_1}  \nonumber \\
&=& \frac{1}{\sqrt{2}} \sum_{ij} \alpha_i \ket{i}_A \ket{\xi_i}_B \ket{i\oplus j}_{A_1} \ket{j}_{B_1},
\end{eqnarray}
where ``$\oplus$'' denotes the addition module 2.

Then she measures the qubit $A_1$ in computational basis $\ket{a}\bra{a},(a=0,1)$, and tell the result $a$ to Bob via a classical communication channel.
The state of $A,B,B_1$ becomes
\begin{eqnarray}
& & \sum_{ij} \alpha_i \ket{i}_A \ket{\xi_i}_B \ket{j}_{B_1} \delta_{a,i\oplus j}\nonumber \\
&=& \sum_{ij} \alpha_i \ket{i}_A \ket{\xi_i}_B \ket{j}_{B_1} \delta_{j,i\oplus a} \nonumber \\
&=& \sum_{i} \alpha_i \ket{i}_A \ket{\xi_i}_B \ket{i\oplus a}_{B_1}.
\end{eqnarray}

\paragraph*{step 2}
If the result $a=0$ Bob does nothing, if $a=1$ Bob performs the operation $X$ on $B_1$, where
\begin{equation}
X =
\left( \begin{array}{cc}
0 & 1\\
1 & 0
\end{array} \right)
\end{equation}
is the first Pauli matrix.
Because $X\ket{i}=\ket{i\oplus 1},(i=0,1)$, the state of $A,B,B_1$ becomes
\begin{equation}
\sum_{i} \alpha_i \ket{i}_A \ket{\xi_i}_B \ket{i}_{B_1}.
\end{equation}

\paragraph*{step 3}
Bob performs the two-qubit operation $U$ on his qubits $B_1$ and $B$.
The state of $A,B,B_1$ becomes
\begin{eqnarray}
& & U^{B_1,B} \sum_{i} \alpha_i \ket{i}_A \ket{\xi_i}_B \ket{i}_{B_1} \nonumber \\
&=& \sum_{i} \alpha_i \ket{i}_A (u_i\ket{\xi_i})_B \ket{i}_{B_1}.
\end{eqnarray}

\paragraph*{step 4}
Bob performs the operation $H$ on $B_1$, where
\begin{equation}
H =
\left( \begin{array}{cc}
1 & 1\\
1 & -1
\end{array} \right)
\end{equation}
is the Hadamard operation.
Because $H\ket{i}=\frac{1}{\sqrt{2}}(\ket{0}+(-1)^i\ket{1})$, the state of $A,B,B_1$ becomes
\begin{eqnarray}
& & \sum_{i} \alpha_i \ket{i}_A (u_i\ket{\xi_i})_B \frac{1}{\sqrt{2}}(\ket{0}+(-1)^i\ket{1})_{B_1}.\nonumber \\
&=& \frac{1}{\sqrt{2}} [ \ket{0}_{B_1} \sum_{i} \alpha_i \ket{i}_A (u_i\ket{\xi_i})_B + \ket{1}_{B_1} \sum_{i} (-1)^i \alpha_i \ket{i}_A (u_i\ket{\xi_i})_B].
\end{eqnarray}

Bob measures $B_1$ in the computational basis $\ket{b}\bra{b},\quad b=0,1$, and tell the result $b$ to Alice via a classical communication channel.
The state of $A,B$ becomes
\begin{equation}
\sum_{i} (-1)^{ib} \alpha_i \ket{i}_A (u_i\ket{\xi_i})_B.
\end{equation}

\paragraph*{step 5}
If $b=0$ Alice does nothing, if $b=1$ Alice performs the operation $Z$ on $A$, where
\begin{equation}
Z =
\left( \begin{array}{cc}
1 & 0\\
0 & -1
\end{array} \right)
\end{equation}
is the third Pauli matrix.
Because $Z\ket{i}=(-1)^i\ket{i}$, the state of $A,B$ becomes
\begin{equation}
\sum_{i} \alpha_i \ket{i}_A (u_i\ket{\xi_i})_B = U^{A,B}\ket{\Psi_0}_{AB}.
\end{equation}

Thus, after these five steps, the diagonal block operation $U$ are determinately implemented on $A,B$ using 1 ebit and 2 cbits.

\begin{figure}[ht]
\begin{center}
\includegraphics[scale=1.0]{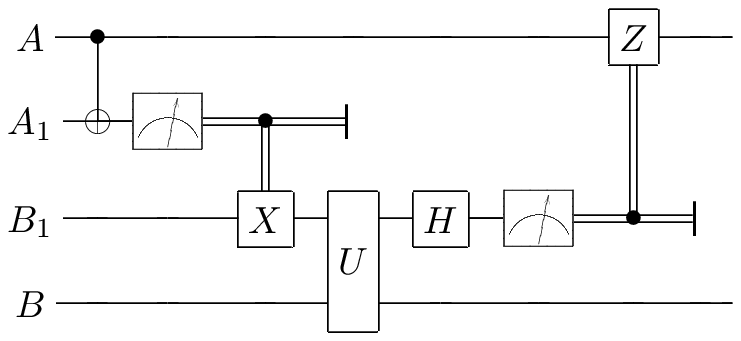}
\end{center}
\vskip -0.1in \caption{ \label{fig1}
Quantum circuit of the protocol for diagonal block operation, 
$A_1$ and $B_1$ is a maximally entangled pair in the state defined by Eq. \ref{e}.}
\end{figure}

This protocol can be expressed as Fig. \ref{fig1}.
It can be found that the protocol is the same as the protocol for the CNOT operation or the control-U operations discussed in Ref. \cite{p00}.
We have just proved that the same procedure can also be used for any diagonal block operation.
This result is nontrivial for the following reasons.

Because the diagonal block operation
\begin{equation}
U=(I\otimes u_0)(\ket{0}\bra{0}\otimes I+\ket{0}\bra{0}\otimes u_0^{\dagger}u_1),
\end{equation}
where $I$ is the identity operation, so if $U$ is completely known, it can also be implemented by performing a control-U operation $\ket{0}\bra{0}\otimes I+\ket{0}\bra{0}\otimes u_0^{\dagger}u_1$ and a single qubit operation $u_0$.
The control-U operation can be implemented using the protocol in Ref. \cite{p00}.
However, if using this method, Bob must construct two new devices to implement these two operations.
These accessorial devices need consume more local resources and may bring a loss of accuracy.
So if Bob already have the device of $U$, it is more economical and more accurate to directly use our protocol.

Furthermore, if Bob has the device of a diagonal block operation, but they do not know the detail of it, then they can not use other operations to simulate it.
However, our protocol is still a choice for them, even in this case.

Investigating the proof of the protocol, it can be found that the protocol is independent on the dimension of the blocks.
So if the blocks in Eq.(\ref{u}) are $2^N\times 2^N$ unitary matrices, i.e., the diagonal block operation operates on $N+1$ qubits --- the first qubit belongs to Alice and the others belong to Bob, they can also locally implement this operation using the same protocol, just replacing the two-qubit operation by the $(N+1)$-qubit operation and replacing the qubit $B$ by these $N$ qubits correspondingly.

\begin{figure}[ht]
\begin{center}
\includegraphics[scale=1.0]{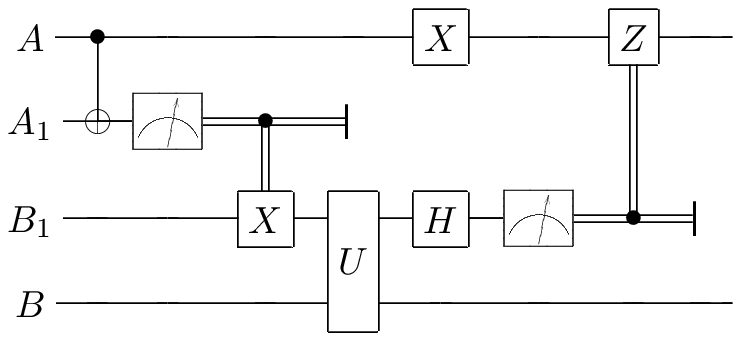}
\end{center}
\vskip -0.1in \caption{ \label{fig2}
Quantum circuit of the protocol for offdiagonal block operation, 
$A_1$ and $B_1$ is a maximally entangled pair in the state defined by Eq. \ref{e}.}
\end{figure}

Consider an offdiagonal block operation, i.e., the operation can be expressed as
\begin{equation}
U^{A,B}=\sum_{i=0}^{1} \ket{i\oplus 1}\bra{i} \otimes u_i,
\end{equation}
where $u_i$s are unitary matrices.
They can locally implement such an operation using the same protocol, except that Alice need first perform an $X$ operation on $A$ in step 5.
This accessorial operation is a local operation at Alice's place, so it is commutable with Bob's local operations in step 2-4.
Thus, Alice can perform this operation at anytime after step 1 and before step 5.
The protocol for offdiagonal block operations can be expressed as Fig. \ref{fig2}.
The validity of it can be proved similarly.

\section{bipartite multiqubit } \label{pnm}
Consider the diagonal block operation in Eq. (\ref{u}). 
If $u_0=c_0 I$ and $u_1=c_1 I$, then $U=(c_0 \ket{0}\bra{0}+ c_1 \ket{1}\bra{1})\otimes I$.
Thus, $U$ is actually a one-qubit diagonal operation, and the protocol in Sec. \ref{p2} is actually a protocol to remotely implement a diagonal operation from Bob to Alice.
In fact, it is just the HPV protocol proposed in Ref. \cite{hpv02}.
The only difference between the one-qubit diagonal operations and the two-qubit diagonal block operations is that the arbitrary coefficients are replaced by arbitrary unitary matrices, and the only difference between these two protocols is that the one-qubit diagonal operation is replaced by the two-qubit diagonal block operation.
In general, intuitively, if there is a protocol for the remote implementation of ``any'' operation that can be expressed as
\begin{equation}
U^{A}=\sum_i c_i K_i,
\end{equation}
where $K_i$s are certain matrices which can be regarded as the characteristic of the restricted set of the protocol, and $c_i$s are arbitrary coefficients,
then the protocol may also be used to locally implement ``any'' block operation that can be expressed as
\begin{equation}
U^{A,B}=\sum_i K_i \otimes u_i,
\end{equation}
where $u_i$s are arbitrary matrices, just replacing $U^{A}$ by $U^{A,B}$.
Because of the linearity of quantum operations, we can expect the validity of this proposition.

Ref. \cite{wang06} generalized the HPV protocol to a protocol for the remote implementation of $N$-qubit operations that can be expressed as
\begin{equation}
U=\sum_{i=0}^{2^N-1} c_i \ket{p_i(x),D}\bra{i,D},
\end{equation}
where $D$ indicates the decimal system, i.e., $\ket{0,D}=\ket{00\cdots0}$, $\ket{1,D}=\ket{00\cdots1}$, $\ket{2^N-1,D}=\ket{11\cdots1}$, etc.
And, 
\begin{equation}
p(x)=\{p_0(x),p_1(x),\cdots,p_{2^N-1}(x)\},
\end{equation}
is a certain permutation of the list $\{0,1,\cdots,2^N-1\}$, where $x=1,2,\cdots,2^N!$ labels all of the $2^N!$ permutations.
We can similarly generalize the protocol in Sec. \ref{p2}.

Alice and Bob need locally implement an $(N+M)$-qubit operation
\begin{equation}
U=\sum_{i=0}^{2^N-1} \ket{p_i(x),D}\bra{i,D} \otimes u_i,
\end{equation}
where $u_i$s are arbitrary $2^M \times 2^M$ unitary matrices.
Bob has the device of $U$.
Alice has the anterior $N$ qubits named $Y_1,Y_2,\cdots,Y_N$, and Bob has the posterior $M$ qubits named $Z_1,Z_2,\cdots,Z_M$.
They share $N$ maximally entangled pairs
\begin{equation}
\ket{\Phi}_{A_iB_i}=\frac{1}{\sqrt{2}}(\ket{00}+\ket{11})_{A_iB_i},\qquad (i=1,2,\cdots,N).
\end{equation}
Alice has the qubits $A_i$s, and Bob has the qubits $B_i$s.
The protocol can be expressed as the following steps.

\paragraph*{step 1}
Alice performs $CNOT^{Y_i,A_i}$ on every $Y_i,A_i$ respectively.
Then she measures every $A_i$ in computational basis respectively and tell Bob the results.

\paragraph*{step 2}
If the measurement result of $A_i$ is $\ket{0}$ Bob does nothing, if the result is $\ket{1}$ Bob performs $X$ on $B_i$ correspondingly.

\paragraph*{step 3}
Bob performs $U$ on $B_i$s and $Z_j$s.

\paragraph*{step 4}
Bob performs an $H$ on every $B_i$ respectively.
Then he measures every $B_i$ in computational basis respectively and tell Alice the results.

\paragraph*{step 5}
Alice performs the permutation operation
\begin{equation}
R(x)=\sum_{i=0}^{2^N-1} \ket{p_i(x),D}\bra{i,D},
\end{equation}
on $Y_i$s.
Then if the measurement result of $B_i$ is $\ket{0}$ Alice does nothing, if the result is $\ket{1}$ Alice performs $Z$ on $Y_i$ correspondingly.

After these 5 steps, Alice and Bob can locally implement the operation $U$ on $Y_1,Y_2,\cdots,Y_N$ and $Z_1,Z_2,\cdots,Z_M$ using $N$ ebits plus $N$ cbits from Alice to Bob plus $N$ cbits from Bob to Alice.
Every $R(x)$ in step 5 can be implemented using two-qubit opearation CNOT and single-qubit operation $X$ \cite{wang06}, so it is not too difficult to implement it.

The validity of this protocol can be proved using the methods similar to the appendix of Ref. \cite{my07}.
The full proof can be found in Appendix \ref{prove}. 
\section{multiparty} \label{p3}
In the protocol in Sec. \ref{pnm}, all of the operations performed by Alice are local to a certain qubit pair $Y_i$ and $A_i$s, except for the permutation operation $R(x)$.
So when $R(x)$ is a direct product of single-qubit operations ($I$ or $X$), the protocol can be generalized to the multiparty cases --- one node has the qubits $B_i$s and each of the other nodes has a pair of $Y_i$ and $A_i$.
We only discuss an example of three-party in this section.
Other cases are all similar to it.

Consider a three-qubit diagonal block operation
\begin{equation}
U=
\left( \begin{array}{cccc}
u_{00} & & & \\
 & u_{01}& & \\
 & & u_{10}& \\
 & & & u_{11}
\end{array} \right),
\end{equation}
where $u_{ij}$s are $2\times 2$ unitary matrices.
This operation is to be implemented on Alice, Bob, and Charlie's qubits $A,B,C$, and Charlie has the device.
This operation can be expressed as
\begin{equation}
U^{A,B,C}=\sum_{i,j=0}^{1} \ket{i}_A\bra{i} \otimes \ket{j}_B\bra{j} \otimes u_{ij}^C.
\end{equation}
In general, this operation can not be transformed to the three-party control-U operation discussed in Ref. \cite{p00}, but it can be locally implemented using the same method in Ref. \cite{p00}.

\begin{figure}[ht]
\begin{center}
\includegraphics[scale=1.0]{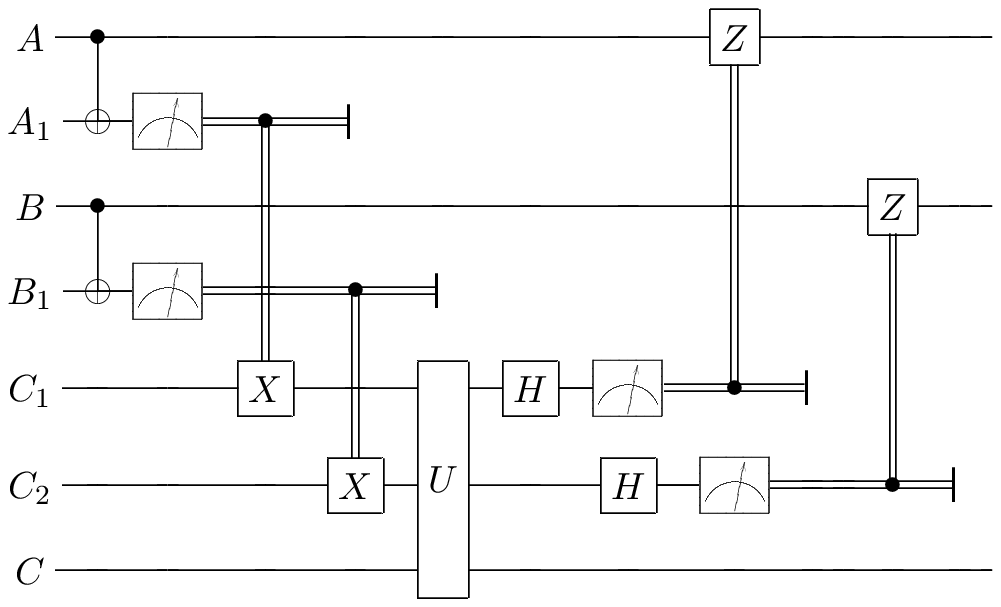}
\end{center}
\vskip -0.1in \caption{ \label{fig3}
Quantum circuit of the protocol for three-part diagonal block operation, 
$A_1C_1$ and $B_1C_2$ are maximally entangled pairs in the state defined by Eq. \ref{e}.}
\end{figure}

Alice and Charlie share a maximally entangled pair $A_1C_1$ as Eq. (\ref{e}).
Bob and Charlie share another maximally entangled pair $B_1C_2$.
They can use the protocol expressed in Fig. \ref{fig3} to locally implement this operation.

\paragraph*{step 1}
Alice performs a CNOT on her qubit $A$ and $A_1$.
Then she measures $A_1$ in computational basis and tell the result to Charlie.

\paragraph*{step 1'}
Bob performs a CNOT on her qubit $B$ and $B_1$.
Then he measures $B_1$ in computational basis and tell the result to Charlie.

Obviously, step 1 and step 1' can be implemented in parallel.

\paragraph*{step 2}
If the measurement result of $A_1$ is $\ket{1}$ Charlie performs an $X$ on $C_1$, and if the result is $\ket{0}$ Charlie does nothing.
If the measurement result of $B_1$ is $\ket{1}$ Charlie performs an $X$ on $C_2$, and if the result is $\ket{0}$ Charlie does nothing.

\paragraph*{step 3}
Charlie performs the three-qubit operation $U$ on $C_1,C_2,C$.

\paragraph*{step 4}
Charlie performs an $H$ on $C_1$ and $C_2$ respectively.
Then he measures $C_1$ and $C_2$ in computational basis and tell the results to Alice and Bob respectively.

\paragraph*{step 5}
If the measurement result of $C_1$ is $\ket{1}$ Alice performs a $Z$ on $A$, and if the result is $\ket{0}$ Alice does nothing.

\paragraph*{step 5'}
If the measurement result of $C_2$ is $\ket{1}$ Bob performs a $Z$ on $B$, and if the result is $\ket{0}$ Bob does nothing.

Thus, the three parters accomplish their task determinately using two ebits and four cbits.
The validity of this protocol can be proved similarly.

\section{conclusion and discussion} \label{con}
In this paper, we proved that any diagonal or offdiagonal block operations can be locally implemented using a similar protocol to the protocol for the CNOT operation, which is discussed in Ref. \cite{p00}. 
The protocol is independent on the dimension of the blocks.
Then we compared the local implementation of nonlocal operations with the remote implementation of local operations, and pointed out a relation between them.
Basing on this comparision, we generalized the protocols in Sec. \ref{p2} to the multiqubit cases in Sec. \ref{pnm}.
Finally, we generalized the protocol to the multiparty cases in Sec. \ref{p3}.

Local implementations of nonlocal operations are important procedures in distributed quantum computation.
These procedures can be implemented using entanglement resources and classical communications.
Entanglement resources are precious in quantum information and quantum computation.
So it is important to implement these procedures economically.
The minimal resources of local implementation of some operations have been found, e.g., the CNOT operation.
However, for most operations, the minimum is still unknown.
Our method would provide some clues for this research.

Recent researches prefer to use Bell states as the entanglement resources.
Nevertheless, other entanglement resources such as GHZ states are also important.
So it is interesting to search appropriate protocols using these resources.

\section*{Acknowledgments}

We acknowledge all the collaborators of our quantum theory group at the Institute for Theoretical Physics of our university. 
This work was funded by the National Natural Science Foundation of China under Grant No. 60573008.

\begin{appendix}
\section{proof of our protocol}\label{prove}
In this appendix, we prove the hybrid protocol proposed in Sec. \ref{pnm}.

The initial state of the qubits $Y_1Y_2\cdots Y_NZ_1Z_2\cdots Z_M$ can always be expressed as
\begin{eqnarray}
& & \ket{\xi}_{Y_1Y_2\cdots Y_NZ_1Z_2\cdots Z_M} \nonumber \\
&=&\sum_{k_1,k_2,\cdots k_{N+M}=0}^1 z_{k_1k_2\cdots k_{N+M}}\ket{k_1k_2\cdots k_{N+M}}\nonumber \\
&=& \sum_{k_1,k_2,\cdots,k_{N}=0}^{1} y_{k_1,k_2,\cdots,k_{N}} |k_1,k_2,\cdots,k_{N}\rangle_{Y_1Y_2\cdots Y_{N}} \otimes |\eta_{k_1,k_2,\cdots,k_{N}}\rangle_{Z_1Z_2\cdots Z_M} \nonumber \\
&=&\sum_{m=0}^{2^N-1} y_{m} |m,D\rangle_{Y_1Y_2\cdots Y_{N}} \otimes |\eta_{m}\rangle_{Z_1Z_2\cdots Z_M},
\end{eqnarray}
where $|\eta_{k_1,k_2,\cdots,k_{N}}\rangle$s or $|\eta_{m}\rangle$s need not be orthogonal each other.
So the initial state of the total system can be expressed as
\begin{eqnarray}
& & \ket{\Psi^{\rm ini}} \nonumber \\
&=& \left(\bigotimes_{m=1}^{N}\ket{\Phi}_{A_mB_m}\right)\otimes\ket{\xi}_{Y_1Y_2\cdots Y_NZ_1Z_2\cdots Z_M} \nonumber \\
&=& \frac{1}{\sqrt{2^N}} \sum_{k_1,k_2,\cdots,k_{N}=0}^{1}  y_{k_1,k_2,\cdots,k_{N}} \bigotimes_{i=1}^{N} \left[\left(\sum_{j=0}^{1} \ket{jj}_{A_iB_i}\right)\ket{k_i}_{Y_i}\right] \otimes |\eta_{k_1,k_2,\cdots,k_{N}}\rangle_{Z_1Z_2\cdots Z_M}. \nonumber \\
\end{eqnarray}

After step 1, the state becomes
\begin{eqnarray}
|\Psi^{1}\rangle &=& \frac{1}{\sqrt{2^N}} \sum_{k_1,k_2,\cdots,k_{N}=0}^{1} y_{k_1,k_2,\cdots,k_{N}}  \nonumber \\
& & \left\{ \bigotimes_{i=1}^{N} \left[(\ket{a_i}_{A_i}\bra{a_i})   CNOT^{Y_i,A_i}\right]\sum_{j=0}^{1} \ket{jj}_{A_iB_i}\ket{k_i}_{Y_i} \right\}
\otimes |\eta_{k_1,k_2,\cdots,k_{N}}\rangle_{Z_1Z_2\cdots Z_M} \nonumber \\
&=& \frac{1}{\sqrt{2^N}} \sum_{k_1,k_2,\cdots,k_{N}=0}^{1} y_{k_1,k_2,\cdots,k_{N}}  \left\{ \bigotimes_{i=1}^{N} \ket{a_i}_{A_i} \ket{k_i\oplus a_i}_{B_i} \ket{k_i}_{Y_i} \right\}
\otimes |\eta_{k_1,k_2,\cdots,k_{N}}\rangle_{Z_1Z_2\cdots Z_M} 
\end{eqnarray}
where $\ket{a_i}$s is the measurement results of $A_i$s.
Every item in the bracket $\{\}$ is calculated similar to Sec. \ref{p2}. 

After step 2, the state of $Y_1Y_2\cdots Y_NZ_1Z_2\cdots Z_MB_1B_2\cdots B_N$ becomes
\begin{eqnarray}
& & \sum_{k_1,k_2,\cdots,k_{N}=0}^{1} y_{k_1,k_2,\cdots,k_{N}}  \left\{ \bigotimes_{i=1}^{N} \ket{k_i}_{B_i} \ket{k_i}_{Y_i} \right\}
\otimes |\eta_{k_1,k_2,\cdots,k_{N}}\rangle_{Z_1Z_2\cdots Z_M} \nonumber \\
&=& \sum_{m=0}^{2^N-1} y_{m} \ket{m,D}_{B_1B_2\cdots B_N} \ket{m,D}_{Y_1Y_2\cdots Y_N} \otimes |\eta_m\rangle_{Z_1Z_2\cdots Z_M}.
\end{eqnarray}
Every item in the bracket $\{\}$ is gotten similar to Sec. \ref{p2}. 

After step 3, the state of $Y_1Y_2\cdots Y_NZ_1Z_2\cdots Z_MB_1B_2\cdots B_N$ becomes
\begin{eqnarray}
|\Psi^{3}\rangle &=& U^{B_1B_2\cdots B_NZ_1Z_2\cdots Z_M} \sum_{m=0}^{2^N-1} y_{m} \ket{m,D}_{B_1B_2\cdots B_N} \ket{m,D}_{Y_1Y_2\cdots Y_N} \otimes |\eta_m\rangle_{Z_1Z_2\cdots Z_M} \nonumber \\
&=& \sum_{m=0}^{2^N-1} y_{m} \ket{p_m(x),D}_{B_1B_2\cdots B_N} \ket{m,D}_{Y_1Y_2\cdots Y_N} \otimes (u_m|\eta_m\rangle)_{Z_1Z_2\cdots Z_M}.
\end{eqnarray}

Denote
\begin{equation}
|p_m(x),D\rangle_{B_1\cdots B_N}=\bigotimes_{i=1}^N |l_m^i(x)\rangle_{B_i}, \qquad (l_m^i(x)=0,1).
\end{equation}
Then,
\begin{eqnarray}
|\Psi^{3}\rangle &=& \sum_{m=0}^{2^N-1} y_{m} |m,D\rangle_{Y_1\cdots Y_N} 
\otimes (u_m|\eta_m\rangle)_{Z_1Z_2\cdots Z_M}
\otimes \bigotimes_{i=1}^N |l_m^i(x)\rangle_{B_i}.
\end{eqnarray}

So, after step 4 the state of $Y_1Y_2\cdots Y_NZ_1Z_2\cdots Z_M$ becomes
\begin{equation}
\sum_{m=0}^{2^N-1} y_{m} (-1)^{\sum_{i=1}^{N} l_m^i(x)b_i } |m,D\rangle_{Y_1\cdots Y_N} 
\otimes (u_m|\eta_m\rangle)_{Z_1Z_2\cdots Z_M},
\end{equation}
Where $b_i$s are the measurement results of $B_i$s.
The coefficient $(-1)^{\sum_{i=1}^{N} l_m^i(x)b_i }$ is gotten similar to Sec. \ref{p2}.

Apparently,
\begin{equation}
R(x) \ket{m,D} = \ket{p_m(x),D}= \bigotimes_{i=1}^N |l_m^i(x)\rangle.
\end{equation}
So after step 5  the state of $Y_1Y_2\cdots Y_NZ_1Z_2\cdots Z_M$ becomes
\begin{eqnarray}
& & \sum_{m=0}^{2^N-1} y_{m} (-1)^{\sum_{i=1}^{N} l_m^i(x)b_i } 
\bigotimes_{i=1}^N [(Z^{b_i})|l_m^i(x)\rangle]_{Y_i}
\otimes (u_m|\eta_m\rangle)_{Z_1Z_2\cdots Z_M}\nonumber \\
&=& \sum_{m=0}^{2^N-1} y_{m} (-1)^{\sum_{i=1}^{N} l_m^i(x)b_i } 
\bigotimes_{i=1}^N (-1)^{l_m^i(x)b_i} |l_m^i(x)\rangle_{Y_i}
\otimes (u_m|\eta_m\rangle)_{Z_1Z_2\cdots Z_M}\nonumber \\
&=& \sum_{m=0}^{2^N-1} y_{m} \bigotimes_{i=1}^N |l_m^i(x)\rangle_{Y_i}
\otimes (u_m|\eta_m\rangle)_{Z_1Z_2\cdots Z_M}\nonumber \\
&=& \sum_{m=0}^{2^N-1} y_{m} |p_m(x),D\rangle_{Y_1Y_2\cdots Y_N}
\otimes (u_m|\eta_m\rangle)_{Z_1Z_2\cdots Z_M}\nonumber \\
&=& U\ket{\xi}_{Y_1Y_2\cdots Y_NZ_1Z_2\cdots Z_M}
\end{eqnarray}

Thus, we accomplish the proof.
\end{appendix}


\begin{thebibliography}{20}
\bibitem{p00}J. Eisert, K. Jacobs, P. Papadopoulos, and M. B. Plenio, Phys. Rev. A {\bf 62}, 052317 (2000)
\bibitem{hpv01}S. F. Huelga, J. A. Vaccaro, A. Chefles, and M. B. Plenio, Phys. Rev. A {\bf 63}, 042303 (2001)
\bibitem{six93} C.H. Bennett, G. Brassard, C. Cr\'epeau, R. Jozsa, A. Peres, and W. K. Wootters, Phys. Rev. Lett. {\bf 70}, 1895 (1993)
\bibitem{clp01}D. Collins, N. Linden, and S. Popescu, Phys. Rev. A {\bf 64}, 032302 (2001).
\bibitem{guo04}Y.-F Huang, X.-F Ren, Y.-S. Zhang, L.-M. Duan, and G.-C Guo, Phys. Rev. Lett. {\bf 93}, 240501 (2004)
\bibitem{d1}W. D$\ddot{u}$r and J. I. Cirac, Phys. Rev. A {\bf 64}, 012317 (2001).
\bibitem{d2}W. D$\ddot{u}$r, G. Vidal, J. I. Cirac, N. Linden, and S. Popescu, Phys. Rev. Lett. {\bf 87}, 137901 (2001).
\bibitem{d3}W. D$\ddot{u}$r, G. Vidal, and J. I. Cirac, Phys. Rev. Lett. {\bf 89} 057901 (2002)
\bibitem{guo05}Yong-Sheng Zhang, Ming-Yong Ye, and Guang-Can Guo, Phys. Rev. A {\bf 71} 062331 (2005)

\bibitem{hpv02}S. F. Huelga, M. B. Plenio, and J. A. Vaccaro, Phys. Rev. A {\bf 65}, 042316 (2002)
\bibitem{wang06}A. M. Wang, Phys. Rev. A {\bf 74}, 032317 (2006)
\bibitem{my07}N. B. Zhao, A. M. Wang, Phys. Rev. A {\bf 76}, 062317 (2007)
\end{thebibliography}
\end{document}